\documentclass[%
 reprint,
 amsmath,amssymb,
 aps,
]{revtex4-1}

\usepackage{graphicx}
\usepackage{dcolumn}
\usepackage{bm}

\begin{document}

\title{Intermittent and continuous flows in granular piles: effects of controlling the feeding height}

\author{L. Alonso-Llanes} \author{L. Dom\'{i}nguez-Rubio}\author{E. Mart\'{i}nez} \author{E. Altshuler}\thanks{corresponding author: ealtshuler@fisica.uh.cu}%
\affiliation{Group of Complex Systems and Statistical Physics, Physics Faculty, University of Havana, 10400 Havana, Cuba.}

\begin{abstract}
Using a specially designed experimental set up, we have studied the so-called continuous to intermittent flow transition in sand piles confined in a Hele-Shaw cell where the deposition height of the sand can be controlled. Through systematic measurements varying the height and the input flow, we have established how the size of the pile at which the transition takes place depends on the two parameters studied. The results obtained allows to explain, at least semi-quantitatively, the observations commonly reported in the literature, carried out in experiments where the deposition height is not controlled.

\begin{description}
\item[PACS numbers]
Hele-Shaw flows, 47.15.gp; Avalanches (granular systems), 45.70.Ht; Avalanches, phase transitions in, 64.60.av; Granular systems, classical mechanics of, 45.70.-n. 
\end{description}
\end{abstract}

\maketitle

Granular media are relevant to many human endeavors: for example, they play a central role in the construction, food and pharmaceutical industries, and also as an important component of the natural environment \cite{andreoti, tejchman, antony}. During the last years granular matter has been increasingly studied from the fundamental point of view by physicists. It has been used, for example, to establish analogies that allow to understand certain phenomena in other areas of physics and engineering, ranging from superconducting avalanches to urban traffic \cite{alt2004,alt2001}.

Granular piles –that we will generically call sandpiles– have been used as a model for segreggation phenomena in geophysical scenarios, and also to illustrate the idea of self-organized criticality \cite{andreoti, aranson, alt2003, etien2007,alt2008}. A particularly attractive configuration of sandpiles is the Hele-Shaw cell: a pile of grain is grown confined between two vertical plates resting on an horizontal surface and separated by a distance $w$, where grains are poured from above, near a third vertical wall also of width $w$ (see Fig. 1) \cite{alt2008, grasselli2000, grasselli97}. The height from which the grains are dropped to feed the pile has been rarely controlled, and only by hand \cite{grasselli2000, grasselli97}. Just recently an automatic system has been designed to fully control the dropping height \cite{leo}.

A well known feature of surface flows in granular piles is the existence of continuous flows (where the grains flows uniformly down the slope within a certain depth from the free surface of the pile) and intermittent flows (where an avalanche suddely rolls down the surface of the pile, and accumulates at its lower edge, allowing a front to grow uphill until a new avalanche starts) \cite{aranson, alt2003, alt2008}]. The transition between the two regimes as the pile grows is, however, poorly understood.

In this communication, we have used the system described in reference \cite{leo} to study the transition from the continuous to the intermittent regime of granular flows on the surface of a quasi-2D pile as a function of the height, $h$, from which the granular matter is fed into the system (see Fig. 1). We put special emphasis in unveiling the relation that exists between “conventional“ experiments –i.e., those where the dropping height is not controlled– and those where it is kept constant in time.

\begin{center}
\begin{figure}
\includegraphics[width=4.25cm,height=4cm]{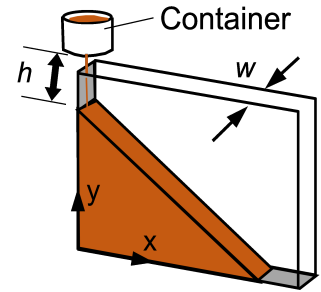}
\caption{\label{fig1} Sketch of the Hele-Shaw cell where the main parameters and coordinates are indicated.}
\end{figure}
\end{center}

In every experiment the behavior of the deposition height, $h$, and the area of the pile were measured as a function of time (the latter allows to compute the input flux, and make sure it is constant along the whole experiment). To obtain the spatial-temporal coordinates of the transition, the evolution in time of a horizontal line of pixels located at a height equal to the half the width of the input flow, from the bottom of the Hele-Shaw cell, was analyzed (see Fig. 2).

\begin{figure}
\includegraphics[width=6.16cm,height=8cm]{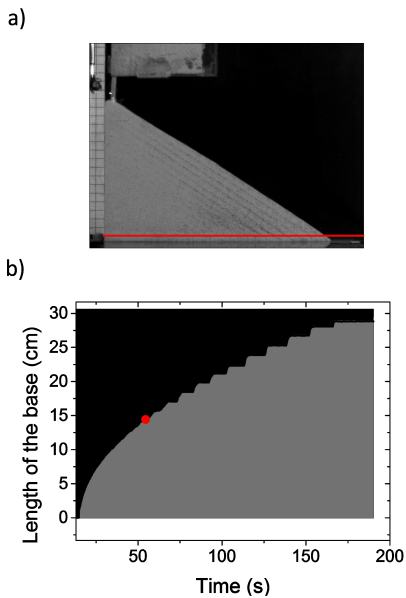}
\caption{\label{fig2} a) Snapshot taken from a video of a typical experiment. The horizontal red line is used to construct the spatial-temporal diagram. b) Spatial-temporal diagram based on the horizontal red line of a). The red dot indicates the coordinate of the continuous to intermittent flow transition.}
\end{figure}

Two types of experiments were performed. In the first group, we kept the feeding height constant in time, as well as the input flux. In the second, the input flux was kept constant, but the dropping height decreased as the pile height increased.  In all experiments, the horizontal size of the pile where the transition from continuous to intermittent flow occurred, $Xc$ , was determined, as shown in Fig. 3.

\begin{figure}
\includegraphics[width=5.69cm,height=8cm]{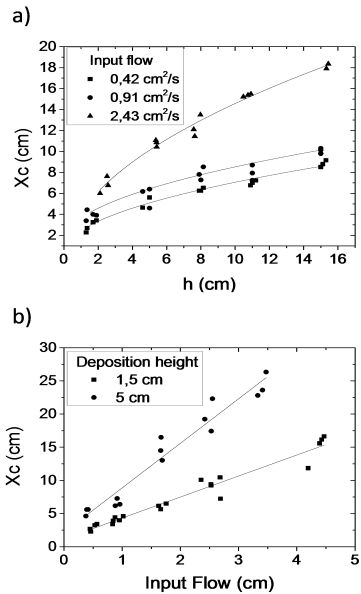}
\caption{\label{fig3} a) Dependence of the transition coordinate $Xc$ as a function of the deposition height for different input flows. The continuous lines follow the law $Xc \propto h^{0.5}$. b) Dependence of the spatial coordinates $Xc$ of the transition as a function of the input flow $F$ for different deposition heights. The continuous lines follow $Xc \propto F$. The data was taken from experiments where h was kept constant. Notice that we repeated at least three times the experiments for each pair of values of deposition height and input flux.}
\end{figure}

Using the obtained temporal coordinates of the transition it is possible, knowing the critical angle for the sand, the input flow and the height of the container at the beginning, to construct a diagram to predict, at least semi-quantitatively, the instant at which the intermittent regime begins in experiments with non-controlled deposition height. It is also possible to find for any experiment with fixed deposition height an equivalent non-controlled experiment in which transition occurs at the same moment, and vice versa, as shown in Fig. 4.

In order to check our predictions, experiments using the same experimental set up were performed, but now without keeping constant the deposition height. The predictions were made applying the mass conservation principle. It was obtained that the deposition height for a non-controlled experiment varies as $ h_{0} - \sqrt{2tFtan\theta_{c}} $, where $h_{0}$ is the initial deposition height, $F$ the input flow and $\theta_{c}$ the critical angle as mentioned above. Figure 4 shows, for one input flow, the temporal coordinates tc of the transition for experiments with different fixed deposition heights, as well as the dependence of the deposition height with time for an experiment where it was non-controlled. The black dots follow a lineal dependence with H that delimitates, from left to right, the end of the continuous phase and the beginning of the intermittent one.

\begin{figure}
\includegraphics[width=8cm,height=5.15cm]{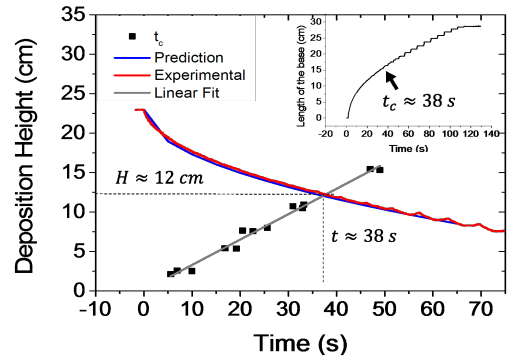}
\caption{\label{fig4} Connection between fixed and variable h experiments. Dots represent the temporal coordinates of the transition measured in our experiments for different values of $h$ where that parameter is constant for each experiment. The red lines are the variation of the deposition height measured for non-controlled experiments –i.e., where h decreases in time as the pile grows.  The blue ones are predictions (see text). Inset: Evolution of the horizontal size of the pile obtained from the spatial-temporal diagram for the non-controlled experiment (red line). The experiments illustrated have an input flux of $2,43 cm^{2}/s$.}
\end{figure}

In summary, we have performed the first systematic study of the transition from the continuous to the intermittent regimes of granular flows on a sand heap, including both conventional experiments, as well as those where the deposition height is controlled to be constant in time. We have demonstrated the relation between the two situations, in such a way that we are able to predict at what size of the pile the transition will take place for a non-controlled deposition height, based on the data taken from experiments with controlled deposition heights. This constitutes a first and necessary step to fully understand the physical nature of the transition.


\begin{thebibliography}{99}

\bibitem{andreoti} B. Andreotti, Y. Forterre and O. Pouliquen, \textit{Granular media} (Cambridge University Press, Cambridge, United Kingdom, 2013.
\bibitem{tejchman} J. Tejchman, \textit{Confined Granular flow in Silos: Experimental and Numerical Investigations} (Springer International Publishing Switzerland, Switzerland, 2013).
\bibitem{antony} S. J. Antony, W. Hoyle and Y. Ding (Editors), \textit{Granular Materials: fundamentals and applications} (The Royal Society of Chemistry, Cambridge, United Kingdom, 2004).
\bibitem{alt2004} E. Altshuler and T. H. Johansen, Rev. Mod. Phys. \textbf{76}, 471 (2004).
\bibitem{alt2001} E. Altshuler, O. Ramos, C. Mart\'{i}nez, L. E. Flores, and C. Noda, Phys. Rev. Lett. \textbf{86}, 5490 (2001).
\bibitem{aranson} I. Aranson and L. S. Tsimring, Rev. Mod. Phys. \textbf{78}, 641 (2006).
\bibitem{alt2003} E. Altshuler, O. Ramos, E. Mart\'{i}nez, A. J. Batista-Leyva, A. Rivera, and K. E. Bassler, Phys. Rev. Lett. \textbf{91}, 014501 (2003).
\bibitem{etien2007} E. Mart\'{i}nez, C. P\'{e}rez-Penichet, O. Sotolongo-Costa, O. Ramos, K. J. M\r{a}l\o{y}, S. Douady, and E. Altshuler, Phys. Rev. E \textbf{75}, 031303 (2007).
\bibitem{alt2008} E. Altshuler, R. Toussaint, E. Mart\'{i}nez, O. Sotolongo-Costa, J. Schmittbuhl, and K. J. M\r{a}l\o{y}, Phys. Rev. E \textbf{77}, 031305 (2008).
\bibitem{grasselli2000} Y. Grasselli, H. J. Herrmann, G. Oron and S. Zapperi, Granular Matter \textbf{2}, 97 (2000).
\bibitem{grasselli97} Y. Grasselli and H. Herrmann, Physica A \textbf{246}, 301 (1997).
\bibitem{leo} L. Dom\'{i}nguez-Rubio, E. Mart\'{i}nez and E. Altshuler, Rev. Cub. F\'{i}sica \textbf{32}, 111 (2015).
\bibitem{dip} B. J\"{a}hne, \textit{Digital Image Processing}, 6th edition (Springer-Verlag Berlin Heidelberg, Germany 2005.

\end{thebibliography}
\end{document}